%% file: bloomRF.tex
\documentclass[sigconf,10pt]{acmart}

\usepackage{graphicx}
\usepackage{subfig}
\usepackage{balance}  
\usepackage{enumitem}
\usepackage{tikz}
\usepackage{amsmath}
\usepackage{stfloats}
\usepackage{latexsym}
\usepackage{url}
\usepackage{xcolor}
\usepackage{tabularx}
\usepackage{booktabs}
\usepackage[ruled,vlined,linesnumbered,noresetcount]{algorithm2e}
\usepackage{varwidth}

\usepackage{mathtools}

\DeclarePairedDelimiter\floor{\lfloor}{\rfloor}

\setcopyright{none}
\acmConference[arXiv online manuscript]{}{2020}

\begin{document}

\title{{bloomRF}: On Performing Range-Queries with Bloom-Filters based on  Piecewise-Monotone Hash Functions and Dyadic Trace-Trees}


\author{Christian Riegger, Arthur Bernhardt, Bernhard Mößner, Ilia Petrov}
 \affiliation{
 	\institution{Data Management Lab, Reutlingen University}
 }	
\email{firstname.surname@reutlingen-university.de}

\newcommand{\bloomRFtitle}{\textsf{\textbf{\lowercase{bloom}RF}}}
\newcommand{\bloomRFcaption}{\textsf{\textbf{bloomRF}}}
\newcommand{\bloomRF}{{\sf \lowercase{bloom}RF}}
\newcommand{\bloomRFparam}[1]{{\sf \lowercase{bloom}RF(#1)}}
\newcommand{\BFs}{Bloom-Filters}
\newcommand{\BF}{Bloom-Filter}
\renewcommand{\shorttitle}{\bloomRF{}}
\renewcommand{\shortauthors}{Mößner, Riegger, Bernhardt, Petrov}
\newcommand{\tn}{tn}
\newcommand{\tp}{tp}
\newcommand{\fp}{f\!p}
\newcommand{\fpr}{f\!pr}
\newcommand{\pot}{pot}

\begin{abstract}
We introduce \bloomRF{} as a unified method for approximate membership testing that supports both \textit{point-} and \emph{range-queries} on a single data structure. \bloomRF{} extends \BFs{} with range query support  and may replace them. The core idea is to employ a dyadic interval scheme to determine the set of dyadic intervals covering a data point, which are then encoded and inserted.
\bloomRF{} introduces \emph{Dyadic Trace-Trees} as novel data structure that represents those covering intervals implicitly. A  \emph{Trace-Tree encoding scheme} represents the set of covering intervals efficiently, in a compact bit representation. Furthermore, \bloomRF{} introduces novel piecewise-monotone hash functions that are locally order-preserving and thus support range querying. We present an efficient membership computation method for \emph{range-queries}. Although, \bloomRF{} is designed for integers it also supports string and floating-point data types. It can also handle multiple attributes and serve as \emph{multi-attribute filter}. 

We evaluate \bloomRF{} in RocksDB and in a standalone library. \bloomRF{} is more efficient and outperforms existing point-range-filters by up to  4$\times$ across a range of settings. 
\end{abstract}

\pagestyle{plain}
\maketitle


\input{sections/01_intro}

\input{sections/02_motivation}

\input{sections/03_hash_enc}

\input{sections/05_operations}

\input{sections/09_optimizations}
\input{sections/07_parameter}

\input{sections/06_datatypes}
\input{sections/10_evaluation}

\input{sections/11_relWork}

\input{sections/12_conclusion}


\balance

\newpage
\bibliographystyle{abbrv}
\bibliography{references}

\end{document}

%% file: sections/01_intro.tex

\section{Introduction}
\label{sec:intro}
Modern data sets are large and grow at increasing rates \cite{Gray:SienceExponentialWorld:Nature:2006}. To process them data-intensive systems frequently need to perform massive full scans that incur significant performance and resource consumption penalties. While indices may reduce the scan pressure, they are not always effective due to size concerns, search predicate selectivity, or due to the high construction/maintenance costs and workload compatibility.

\emph{Filters} are a class of approximate data structures that may effectively complement the workhorse data structures or system operations to reduce scans. \emph{\BFs{}} \cite{Bloom:BF:CACM:1970}, are prominent representatives that can perform fast approximate set membership tests. They are efficient, compact and maintain a small and fixed-size bit array. \BFs{} allow \emph{no false negatives}, while false positives are possible, however the \emph{false positive rate (FPR)} can be controlled by parameters such as bits per key or number of hash functions. If the \BF{} returns true, the search key may be present or not and the system needs to verify that by performing an expensive scan or an index-lookup.
\BFs{} only support point-lookups, i.e. is key 4711 in the dataset.  \emph{They are inadequate for range-queries, due to the lack of practicable monotone hash functions.} Indeed, the hash-functions in \BFs{} seek to scatter the data in a perfectly random manner across the bit array, thus precluding range querying.
\begin{figure}[!t]
	\begin{center}
		\includegraphics[width=\columnwidth]{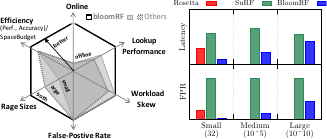}
		\caption{\bloomRFcaption{} extends \BFs{} with range queries, is efficient and performs well (RocksDB).}
		\label{fig:head}
	\end{center}
\end{figure}

{\bf State-of-the-Art Overview.} 
Many algorithms and systems necessitate \emph{efficient range filtering} for queries such as: does the dataset contain \emph{keys ranging between 42 and 4711}. Classical approaches such as Prefix \BFs{} or Min/ Max indices (fence pointers, ZoneMaps in Neteeza or Block-Range Index in PostgreSQL) can perform  range-filtering, but are impractical for point queries and result in a higher FPR.

Rosetta \cite{Dayan:Rosetta:SIGMOD:2020}, SuRF \cite{Zhang:SURF:SIGMOD:2018} and ARF \cite{Alexiou:ARF:VLDB:2013} are some recent proposals that can handle point- and range-lookups on a unified data structure and serve as \emph{point-range filters} (PRF). ARF \cite{Alexiou:ARF:VLDB:2013} and SuRF \cite{Zhang:SURF:SIGMOD:2018} utilize \emph{tries} and thus partially materialize the index at the cost of extra space. Such techniques result in increased range-filter sizes, that are reduced by trie-truncation techniques or require tedious training/re-optimization. These, in turn, result in an \emph{a posteriori/offline} creation. Rosetta \cite{Dayan:Rosetta:SIGMOD:2020} employs a different approach, where each key is decomposed into a set of prefixes according to a dyadic interval scheme and implicit Segment-Trees \cite{deBerg:SegmentTrees:2008}. The prefixes are maintained in a set of hierarchical \BFs{} and each \BF{} contains equi-length prefixes, while the Segment-Trees are \emph{not} materialized. 

\noindent{\bf Problem 1: Existing point-range-filters are designed either for small-sized or for large query ranges}. 
Existing Point-Range-Filters (PRF) are optimized for handling different query range sizes. While Rosetta \cite{Dayan:Rosetta:SIGMOD:2020} excels for small query-ranges $[2^{1}-2^{6}]$, SuRF \cite{Zhang:SURF:SIGMOD:2018} offers excellent FPR for mid- and large- query ranges [$2^{37}-2^{38}]$. On the one hand, the trie-truncation techniques (like the ones used in ARF and SuRF) may loose effectiveness as short query ranges may fall in the scope of the truncated suffixes and thus have higher probability of being detected non-empty. On the other hand, if mid-sized query ranges are frequent, testing against a hierarchical set of \BFs{}, may increase CPU-costs as it implies probing more \BFs{}. 

\noindent{\bf Problem 2: Existing point-range-filters are offline.}
Existing PRF (such as ARF \cite{Alexiou:ARF:VLDB:2013}, SuRF \cite{Zhang:SURF:SIGMOD:2018}, and Rosetta \cite{Dayan:Rosetta:SIGMOD:2020}) employ powerful optimizations, which however require \emph{a priori} the complete key-/data-set and are therefore constructed \emph{offline}. Hence, the PRF cannot serve range-queries, while data is being simultaneously inserted. 
While, systems such as KV-stores, avoid the issue by constructing the PRF only as the data is persisted, and handle searching the main-memory delta by auxiliary structures, such approach is far from optimal in other settings. 
Alternatively, Prefix-BF and Min/Max filters may be constructed online, but are inadequate for point-querying. Rosetta \cite{Dayan:Rosetta:SIGMOD:2020} may be used online per se, however some of its powerful optimizations are offline. 

\noindent{\bf Problem 3: Existing Point-Range-Filters exhibit non-robust performance across a variety of workload distributions.}
Existing PRF are sensitive to access skew. Rosetta \cite{Dayan:Rosetta:SIGMOD:2020} outperforms SuRF by 2$\times$ on a normally distributed workload as the suffix-truncation techniques yield more prefix-collisions for small ranges. More skewed distributions (e.g. Zipfian) yield more collisions and more LSM-Tree levels need to be range-filtered. Furthermore, it is not always possible to assume upper bounds on the query-range size. While, short-ranges seem reasonable for KV-stores, this does not apply to analytical workloads. 
A unified PRF must exhibit robust performance over a wide range of workloads. 

\noindent{\bf Bloom-Range-Filter} \textsf{\textbf{(bloomRF).}} 
We introduce \bloomRF{} as a unified data structure, supporting approximate \emph{point-} and \emph{range-} membership tests that can effectively substitute existing \BFs{}. The core intuition is to employ a \emph{dyadic interval scheme} to determine all intervals covering a data point/key, and thus allow for range-querying. 
Instead of materializing and persisting all covering dyadic intervals that would otherwise increase the space consumption and reduce  performance, \bloomRF{} employs novel \emph{Dyadic Trace-Trees} (DTT). They are implicit balanced binary trees with a logical root and Trace-Trees as nodes and allow encoding the dyadic interval information for each key compactly. 
The {\bf contributions} of this paper are:
\begin{itemize}[leftmargin=*,noitemsep,nolistsep,nosep]
	\item \bloomRF{} is a unified point-range-filter that can effectively replace existing \BFs{}.
		
	\item \bloomRF{} is designed for \emph{online} operation and can serve queries, while data is being simultaneously inserted.

	\item \bloomRF{} employs a novel \emph{Trace-Tree encoding} scheme to encode $h$ covering dyadic interval levels with a single bit, where $h$ is the height of a Trace-Tree and a configurable parameter. Hence, \bloomRF{} can minimize the amount of extra interval data to be written, thus reducing the size, the number of accesses and improving performance.
	
		\item We introduce novel \emph{piecewise-monotone hash functions (PMHF)} that preserve local order, which is essential for range querying. 

		\item \bloomRF{} can serve small- to large-query ranges with acceptable FPR, and can handle different workloads.
					
		\item Although \bloomRF{} is designed for \emph{integer domains}, it  supports \emph{string} and \emph{floating-point} datatypes. Floating-point numbers are relevant for scientific DBMS or AI.
		
		\item Furthermore, \bloomRF{} can serve as a \emph{multi-attribute} filter, which is especially important for analytical settings.
		
		\item \bloomRF{} is configurable. Its tuning advisor  computes parameter settings to address the memory-FPR tradeoff.

		\item \bloomRF{} outperforms all baselines by 5\% to 4$\times$ across a range of settings (volume, query range sizes). \bloomRF{} is more efficient as it achieves better performance and FPR at lower space budgets (bits per key).	
\end{itemize}

\begin{figure*}[!t]
	\begin{center}
		\includegraphics[width=1.02\textwidth]{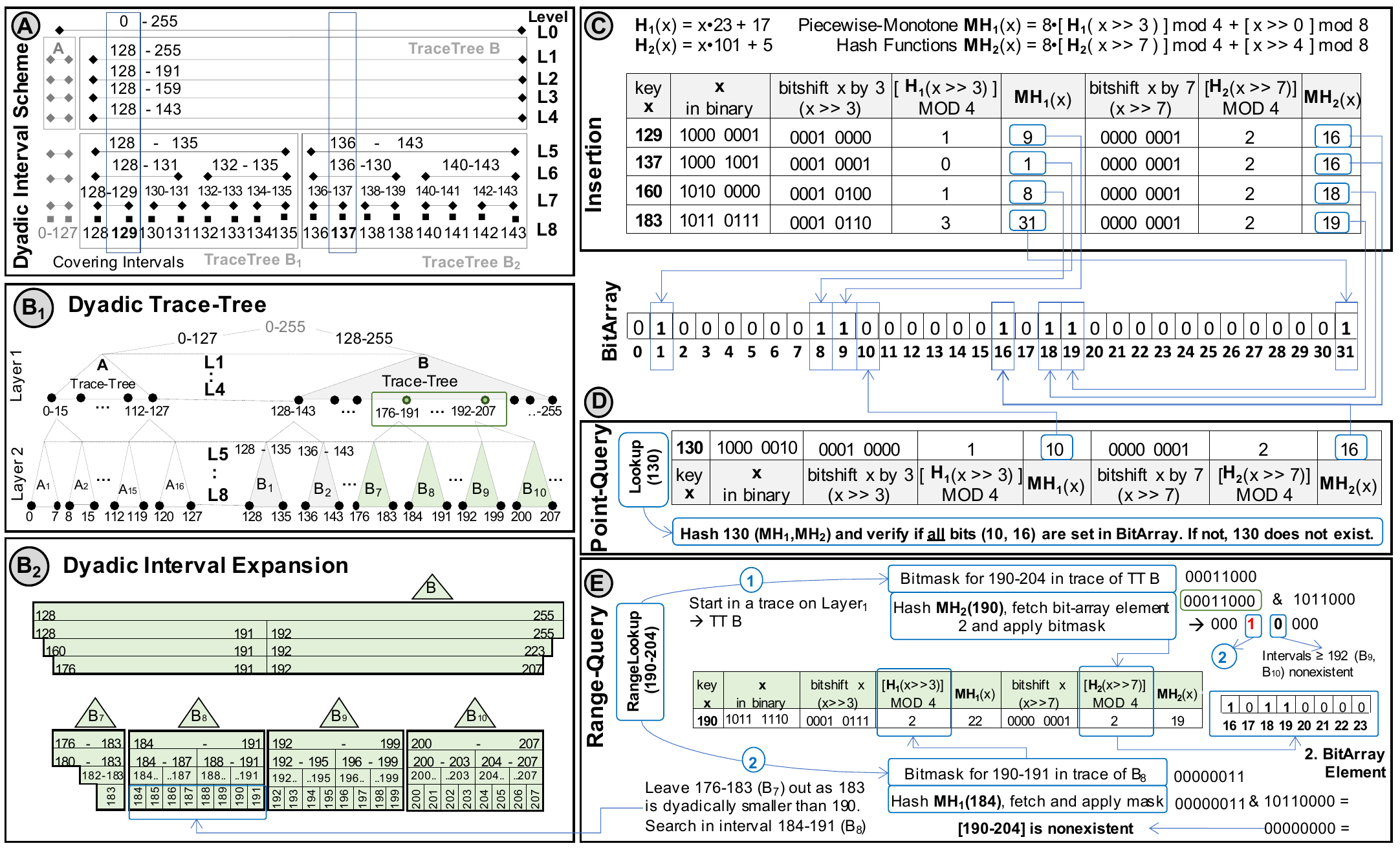}  
		\caption{ \bloomRFcaption{} relies 
		(a) on a \emph{dyadic	 interval scheme}; 
		which is represented as a   
		(b$_{1}$) \emph{Dyadic Trace-Tree} where each data point 
		(b$_{2}$) represents an interval and is encoded with a
		Trace-Tree scheme and  
		(c) \emph{piecewise- monotone hashing functions}. Upon \emph{insertion} they place it in a \emph{bit array}, which is used in 
		(d) \emph{point-} and (e) \emph{range-lookups}.}
	\label{fig:headpict}
	\end{center} 
\end{figure*}

{\bf Outline.} We motivate basic \bloomRF{} with a brief end-to-end example (Sect. \ref{sec:motivation}) and present optimizations in Sect. \ref{sect:optimization}. The hashing and encoding scheme is described in Sect. \ref{sect:hash:enc}, the  range-lookup algorithm is in Sect. \ref{sect:ops}. Sect. \ref{sect:param} describes the \bloomRF{} tuning advisor built atop an FPR model. We discuss the experimental results in Sect. \ref{sect:eval} and conclude in Sect. \ref{sect:conclusions}.

%% file: sections/02_motivation.tex

\section{Overview} 
\label{sec:motivation}
We begin with brief, but complete walk-through of all main aspects of \bloomRF{} (Fig. \ref{fig:headpict}), before describing their details in depth throughout the sections to follow. 

\bloomRF{} is designed to cover the full datatype domain $D$ of size $d$-bits, $|D|=2^{d}$. Conceptually, \bloomRF{} is based on a \emph{Dyadic Interval Scheme} (Fig. \ref{fig:headpict}.a), covering $D$. It comprises $\ell \in \{0,1,\dots, d\}$ dyadic levels (Fig. \ref{fig:headpict}.a) and each dyadic interval on a level $\ell$ spans $2^{d-\ell}$ keys. An excerpt of the dyadic scheme for a one-byte unsigned integer, $|D|\!=\!2^{8}$ is shown in Fig. \ref{fig:headpict}.a.

Interestingly, if the numbers of all $L$ covering intervals were to be inserted as extra $d$ data points for each inserted key, even a standard \BF{} would be able to answer a range query. This can be done by performing an exact dyadic decomposition of the query-interval $I_{Q}$ into at most $2\cdot log_{2}(I_{Q})$ dyadic intervals $I_{d}$ followed by at most $I_{d}$ point lookups. While possible, such solution is impractical as it yields \emph{prohibitively large} filter-sizes.

To this end, we introduce the \emph{Dyadic Trace-Trees -DTT} (Fig. \ref{fig:headpict}.$b_{1}$ and Sect. \ref{sect:hash:enc} for more details), as a novel data structure that drastically reduces the amount of extra data per key. A DTT is an implicit balanced binary tree with a logical root and \emph{Trace-Trees (TT)} as nodes, organized in multiple layers of TTs. Each Trace-Tree encodes the membership in $h$ hierarchical dyadic intervals as a single data point in a single bit. Thus a TT avoids materializing $h-1$ dyadic levels and does so only for the $h^{th}$, where $h$ is configurable. 

A DTT reflects the dyadic interval scheme. Consider key 129 (Fig. \ref{fig:headpict}.a), for instance. It is covered by the following dyadic intervals \{$[$129,129$]$, $[$128,129$]$,$[$128,131$]$, \dots, $[$128,255$]$, $[$0,255$]$\}. If subdivided into groups of h=4 dyadic levels, it results in a DTT with two Trace-Tree layers: on the upper layer TT \emph{A} (not shown in its entirety) and TT \emph{B}, comprising \{$[$128,255$]$,$[$128,191$]$,$[$128,159$]$,$[$128,143$]$\}, i.e. levels $L_{1}$-$L_{4}$, while on the lower we have TT $A_{1}$\dots $A_{16}$ (not depicted)  together with TT $B_{1}$ comprising \{$[$129,129$]$, $[$128,129$]$, $[$128,131$]$, $[$128,135$]$\}, i.e. levels $L_{5}$-$L_{8}$, as well as TT $B_{2}$\dots $B_{16}$ (shown partially). Fig. \ref{fig:headpict}.b$_{2}$ shows a DTT excerpt ($B$, $B_{7}$-$B_{10}$), needed for the range-query 190-204 described below.

Each DTT comes with a set of locally order-preserving, piecewise-monotone hash-functions (\emph{PMHF})-- one for each Trace-Tree layer in a DTT. In the running example we have two: $MH_{1}$ and $MH_{2}$ (Fig. \ref{fig:headpict}.c and Sect. \ref{sect:hash:enc}). $MH_{1}$ focuses the $h\!=\!4$ least-significant bits, while $MH_{2}$ covers the four MSB.

The \emph{insertion} of a key in \bloomRF{} (Fig. \ref{fig:headpict}.c and Sect. \ref{sect:ops} for more details) is analogous to a \BF{} insertion. \emph{PMHF} are applied on the respective $h$ bits, to compute a single bit per layer. Consider, the insertion of \emph{129} (Fig. \ref{fig:headpict}.c).  Applying $MH_{1}$ (TT $B_{1}$), yields setting a bit in the $9^{th}$ bit-array position, while applying $MH_{2}$ on the four MSBs (TT $B$) yields setting the $16^{th}$ bit. Hence, 129 is inserted by setting just two bits. 

A \emph{point-query} in \bloomRF{} resembles the insertion. $MH_{1}$ and $MH_{2}$ are applied on the key (Fig. \ref{fig:headpict}.d and Sect. \ref{sect:ops} for more details) and the computed bits are retrieved. If at least one of the bits is 0 the key is non-existent. 

\emph{Range-Queries} in \bloomRF{} can be performed intuitively, based on the dyadic scheme. The range search (Fig. \ref{fig:headpict}.e and Sect. \ref{sect:ops}) first computes lowest DTT layer, for which the query range \texttt{[190,204]} is in a single trace, i.e Layer 1 and TT \emph{B} in the running example. Next, a bitmask for all intervals covered by the search range in that trace is computed. This is possible since intervals are encoded as bits and since PMHF ensure their order in a TT trace. Thus, the bitmask is \texttt{00011000} since only $[$176, 191$]$, $[$192, 207$]$ cover the query range in the trace of $B$ in Fig. \ref{fig:headpict}.b$_{1}$. We zoom in this part of the DTT in Fig. \ref{fig:headpict}.b$_{2}$; remember that the positions in a TT trace (e.g. of $B$) are actually intervals. The left search key is hashed with $MH_{2}(190)$, the PMHF on layer 1 and the bit array element corresponding to $B$ is fetched. To check if the child TTs (on layer 2) do not include the interval, a bitwise \emph{AND} with the bitmask is performed, yielding \texttt{00010000}. The \texttt{1} indicates that subintervals $[$176,183$]$, $[$184,191$]$ may be non-empty, i.e. child TTs $B_{7}$ and $B_{8}$ need to be checked, while the succeeding \texttt{0} indicates that intervals $\geq 192$ are empty. Since $[$176,183$]$ is outside the search range it is also discarded, and only the range $[$184,191$]$ in $B_{8}$ is checked. Next, we repeat the process on layer 2, by first computing the bitmask for \texttt{[190,191]} for TT $B_{8}$, i.e. \texttt{00000011}, hashing $MH_{1}(184)$ and applying the bitmask. Since the result is \texttt{0} there are no members in $B_{8}$ and the search range \texttt{[190,204]} is empty.

%% file: sections/03_hash_enc.tex

\begin{figure}[!h]
	\centering
  \includegraphics[width=.9\linewidth]{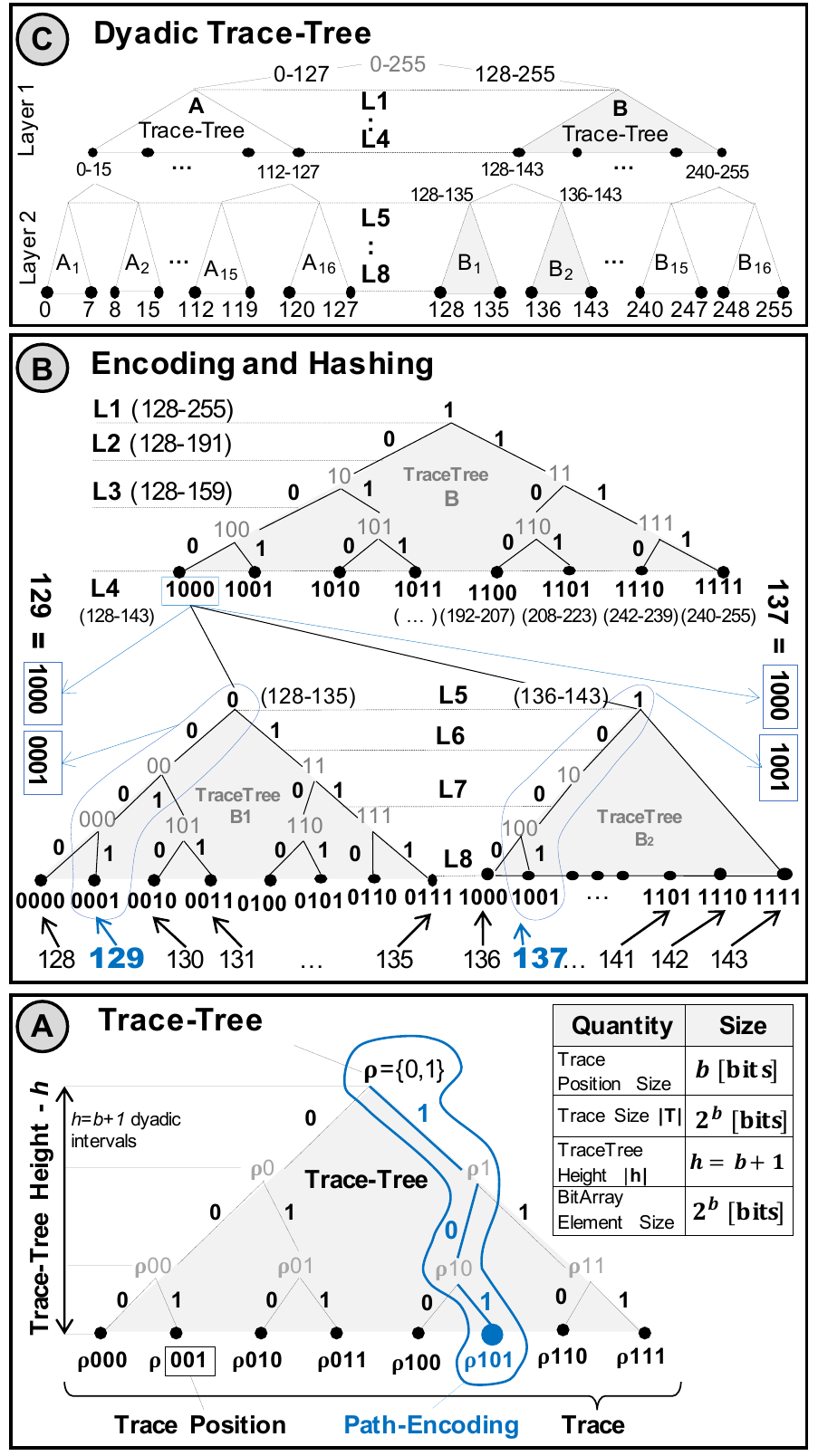}
  \caption{Organization of a (a) trace-tree and path-encoding; (b) example of the trace-tree representation of 129 and 137 in (c) an excerpt of the DTT.}
  \label{fig:TT} 	
\end{figure}

\begin{table}[b]
	\footnotesize 
    \centering
    \begin{tabularx}{\columnwidth}{@{}lXl@{}}
        \toprule
        \midrule
		{\bf b} & number of bits encoding a trace position, $b=log_{2}(|T|)$ \\  		{\bf |T|} & Trace size in bits -- $2^{b}$ \\
		{\bf D, |D|} & attribute domain and size in elements $ |D|=2^{d}$, $2^{32}$ for UINT32\\		
		{\bf (D)TT} & (Dyadic) Trace-Tree\\
		{\bf |H|} & Height of a DTT in layers of TT, 
		$|H| = \floor*{  LOG_{(2\cdot |T|)} \left( \frac{|D|}{|T|} \right) } + 1$\\
		{\bf h} & Height of a TT in dyadic levels, $h = b+1$\\
		{\bf MH$_{i}$} & piecewise-monotone hash function $MH_{i}(x) ,i\in[1,|H|]$\\
		{\bf $x^\ell$} & dyadic intervals on level $\ell \in \{0,1,\dots, d\}$, $x \in D$\\
		{\bf $\mathcal T_y^i$} & dyadic TT on layer $i \in \{0,1,\dots,L\}$, $y \in D$\\ 
		{\bf $\ell_i$} & lowest level of layer $i$, i.e. $\ell_i = i\cdot h + d \! \! \mod h$\\ 
        \bottomrule
    \end{tabularx}
    	\caption{Terms and Abbreviations}	
    	\label{tab:abbr}
\end{table}

\section{Encoding and Hashing} 
\label{sect:hash:enc}

As stated at the beginning of Sect. \ref{sec:motivation} even standard \BFs{} can answer a range query. However, \emph{the filter size becomes prohibitively large.} For 5 million, uniformly distributed keys over $|D|=2^{64}$, 48 bits/key (30MB) are needed to achieve FPR of 0.18 (18\%) and 400 bits/key for FPR 0.001 (\BF{} uses Murmur hash-functions.) To address the size issues we introduce novel encoding and hashing schemes.

\noindent{\bf Dyadic Interval Scheme (DIS).} 
Let the set of binary numbers represented with $d$ bits be: $
  D\!=\!
  \{
    x\!=\!
   (x_{d-1}, \ldots x_1, x_0)_2
    \!=\!
    \sum_{i=0}^{d-1} x_i2^i \, | 
    \, x_i \in \{0,1\}
  \}
$.
The dyadic intervals $x^\ell$ are defined by ranging over the last $\ell$ bits as
$ x^\ell \!=\! [0,2^{d-\ell}-1] + \sum_{i=d-\ell}^{d-1} x_i2^i,$
$\ell \!\in\! \{0,1,\ldots,d\}.$
The canonical \emph{dyadic interval scheme} is the set of all these intervals
$
  \mathcal I \!=\!
  \bigcup_{\ell=0}^d \mathcal I^\ell
  ,
  \mathcal I^\ell \!=\!
  \{
    x^\ell \, | \, x \in D
  \}
.
$
It contains $d\!+\!1$ dyadic levels $\mathcal \!I^\ell$, $\ell \!\in \!\{0,1,\ldots,d\}$, where every level $\ell$ comprises $2^\ell$ intervals, each containing $2^{d-\ell}$ elements.


\noindent{\bf Encoding.} \emph{Trace-Trees} (TT) are the foundation of the interval encoding scheme behind \bloomRF{}. 
\emph{Each TT encodes the membership in $h\!=\!b\!+\!1$ hierarchical dyadic intervals as a single data point in a single bit.} Technically, a TT (Fig. \ref{fig:TT}.a) is a binary tree, with bit-sized nodes. It encodes a distinct set of dyadic intervals on $h$ hierarchical dyadic levels, i.e. TT \emph{height}. Each ``bit-node'' corresponds to a dyadic interval, and has exactly two child ``bit-nodes'' according to DIS. The root of a TT ($\rho$) corresponds to the highest dyadic interval. We call the leaves of a TT a \emph{trace} that comprises an \emph{ordered set} $2^{b}$ elements called \emph{trace positions}.  

\emph{Path-encoding} (Fig. \ref{fig:TT}.a) is a core characteristic of a TT: \emph{each trace position encodes implicitly the path to the root $\rho$}. Consider position $\rho101$: the binary representation of the $5^{th}$ trace position is $101$, which in turn is the encoding of the binary path to the root $\rho$. Therefore, TT are implicit as only the trace is \emph{materialized}, but not the upper levels in a TT. 

Each trace corresponds to a bit-array element, where \emph{every trace position is represented by a single bit} at the offset indicated by the trace position number. (Individual bit-array elements are \emph{bit-overlapped} in the bit-array.) Hence, the membership in $h$ dyadic intervals corresponds to a single trace position in a trace tree and is encoded by a single bit.
Consider, for instance,  $129$ in TT $B_{1}$ (Fig. \ref{fig:TT}.b)
Assuming a byte-sized trace and bit-array element, there are $8\!=\!2^{b}$ trace positions and thus $b\!=\!3$. The height of the TT $B_{1}$ is $h\!=\!b\!+\!1\!=\!4$ dyadic interval levels, i.e $\{$[$129,129$]$, $[$128,129$]$, $[$128,131$]$, $[$128,135$]$\}$. 129 corresponds to the second trace position in the trace and thus to the second bit in the bit-array element. Its binary code is $001$, which is the exact binary path to the root of $B_{1}$. 

\textsf{Insights:} 
(1) TT are \emph{implicit} as only the \emph{trace} positions of a TT are materialized, but not the upper TT levels; 
(2) By setting a single bit for the trace position a TT encodes the whole path to the root $\rho$ and therefore the membership in all $h$ hierarchical dyadic intervals;
(3) Trace positions are \emph{implicitly ordered} within a trace and encoded as a single bit, thus traces and their bit-array elements are order-preserving. 
 (1) and (2) translate into low space consumption, and less probes, while (3) allows for range-querying.

\begin{figure}[!b]
	\begin{center}
		\includegraphics[width=.9\columnwidth]{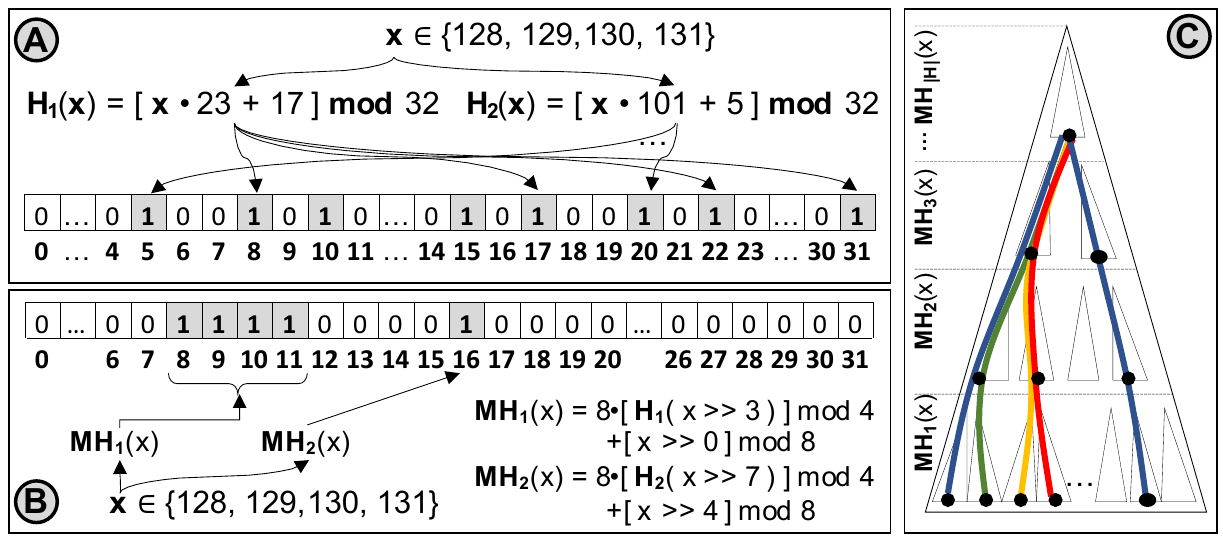}
		\caption{
		While
		(a) hashing in \BFs{} improves random distribution,  
		(b) piecewise-monotone hashing in \bloomRFcaption{} preserves partial order (within a trace) assisting range-querying, and is
		(c) hierarchical in DTT.
		}
		\label{fig:hashfunc}
	\end{center}
\end{figure}

\noindent{\bf Dyadic Trace Trees (DTT)} are implicit balanced binary trees with a logical root and TTs as nodes.
A DTT $\mathcal T$ comprises $L$ logical layers of TTs (Fig. \ref{fig:headpict}.$b_{1}$).
We assume a single uniform trace size $|T|=2^b$, $h=b+1$,
for the whole DTT for the time being. We will relax this assumption in Sect. \ref{sect:optimization}.
The height $|H|$ of a DTT (Fig. \ref{fig:headpict}.$b_{1}$) is the number of layers $L$, i.e., 
$|H| \!=\! \floor*{  LOG_{(2\cdot |T|)} \left( \frac{|D|}{|T|} \right) } \!+\! 1 \!=\!
L \!=\! \floor*{ d/h }$, $|D|\!=\!2^{d}$.
The lowest level of layer $i \in \{0,1,\ldots,L\}$ is then $\ell_i = i\cdot h + d \! \! \mod h$.
More formally: a \text{TT} $\mathcal T_y^i$ is defined as
$
  \mathcal T_y^i
  =
  \{
    x^\ell \subseteq y^{\ell_{i-1}+1}
    \, | \,
    x \in D, \ell \leq \ell_i
  \}
  ,
  i \in \{1,2,\ldots,L\},
  y \in D
  .
$
The set of dyadic intervals $x^i$ on the bottom of a TT
$
  T_y^i
  =
  \mathcal T_y^i
  \cap
  \mathcal I^{\ell_i}
  ,
  i \in \{1,2,\ldots,L\},
  y \in D,
$
constitutes its trace.
Thus $\mathcal T = \bigcup_{i=1}^L \mathcal T^i$,
where a DTT layer $\mathcal T^i$ is the set of all TTs on the same level 
$
  \mathcal T^i
  =
  \{
     \mathcal T_y^i \, | \, y \in D
  \}
  .
$
There are
$2^{\ell_{i-1}+1}$ TTs on layer $i$
and
$TT_{MAX}\!=\!\sum_{i=1}^L 2^{\ell_{i-1}+1}$
.


\noindent{\bf Piecewise-Monotone Hashing.} Another key characteristic of \bloomRF{} is its use of novel piecewise-monotone hash functions (PMHF). While, standard \BFs{} employ a set of independent hash functions aiming to distribute keys in a perfectly uniform manner, PMHF in \bloomRF{} seek to preserve partial order (Fig. \ref{fig:hashfunc}) and hash hierarchically. 

Hashing in \bloomRF{} is \emph{hierarchical} with a separate PMHF on each TT layer in a DTT. This has three noticeable aspects. \emph{Firstly}, there is a family of PMHF $MH_{i}, i\in[1,|H|]$. Each takes the most significant $\ell_i$ bits of data point \emph{x} and computes the bit-array position for the TT trace-position on DTT layer \emph{i}:
\[
\small
\begin{array}{c}
	MH_{i}(x) = \Big[\Big\{ \big[ H_{i}\big(x \!>\!> \!(i\cdot(b\!+\!1)\!-\!1\big)\big]
	 \,mod \,(m >\!> b)  \Big\} <\!< b \Big] \\
	+ [ x >\!> (i-1)\cdot(b+1) ] \, mod \, (2^{b})
\end{array}
\]
\normalsize

\noindent where $H_{i}(x)\!=\!Prime_{i,1}\cdot x+Prime_{i,2}$ $i\in[1,|H|]$, \emph{m} is the number of bits in the bit-array and \texttt{>>} is a bit-shift. We make no special assumption with respect to the PMHF except that they align to  trace sizes. 

\emph{Secondly}, a bit-portion of data point $x$ is hashed by each  $MH_{i}$. Therefore, $x$ has a vertical counterpart on each of the overlying TT layers $i$ in a DTT. 
For example, the DTT in Fig. \ref{fig:headpict}.b$_{1}$ has two TT layers and therefore two PMHF $MH_{i} \in \{MH_{1}, MH_{2} \}$. A data point, e.g. 129 is hashed with $MH_{1}$ for layer 2 and with $MH_{2}$ for layer 1 and has therefore a representative on each of the layers (Fig. \ref{fig:headpict}.c and \ref{fig:hashfunc}), i.e. bit 9 (layer 1) and bit 16 (layer 2). 

Because of the logarithmic nature of a DTT, the vertical counterparts may overlap above a certain layer, as a TT on every non-leaf DTT layer may have $2\cdot|T|$ child TTs on the underlying layer. Consider, for instance, data points 129 and 137 (Fig. \ref{fig:TT}.b), they hash to the second trace positions in TT $B_{1}$ and $B_{2}$ and bits 9 and 1 through $MH_{1}$, but both hash to the first trace position in TT $B$ and bit 16 through $MH_{2}$. Thus lower DTT layers tend to be sparser, while upper are denser and the highest ones saturate (uniform distribution).

\emph{Thirdly}, hierarchical hashing allows for \emph{error-correction}, which is especially important given the overlapped/lossy bit-array. Given a key \emph{x}, if the bit-position of $MH_{k}(x)$ is set, but the bit-position for a higher layer $MH_{j}(x)$ is not set, than it is impossible that \emph{x} exists, because when key \emph{x} is inserted all vertical representatives of $MH_{i}(x)$ (Fig. \ref{fig:hashfunc}.c) must be set. Consider probing \emph{99} and the example of Fig. \ref{fig:hashfunc}. $MH_{1}(99)$ computes position \emph{11}, which is set, but  $MH_{2}(99)$ computes position \emph{14}, thus \emph{99} does not exist. This property is essential for handling skew in \bloomRF{} and for range-querying.

%% file: sections/05_operations.tex

\section{\bloomRFtitle{} Operations}
\label{sect:ops}

\noindent{\bf Range-Lookup.} 
We now describe how \bloomRF{} performs range queries based on Algorithm \ref{alg:rangeq} and on the example in Fig. \ref{fig:headpict}.e. 
Starting from the top DTT layer, indicated by the height (${\bf |H|}$, line \ref{alg:rq:line:init}) we append all dyadic intervals (\emph{check}) intersecting the search range \texttt{[l\_key,r\_key]} to a vector of \emph{intervals} to be inspected (line \ref{alg:rq:line:start:check}). The initial layer may be the exact layer for given a space management strategy (Sec. \ref{sect:optimization} and Fig. \ref{fig:space:allocation}), or the top, non-saturated layer in basic \bloomRF{}. For large ranges several adjacent dyadic \emph{check} intervals may be appended. Assume that the search interval falls in exactly one TT as in the example (Fig. \ref{fig:headpict}.b$_{1}$ and Fig. \ref{fig:headpict}.e).

\begin{algorithm}[t]
\caption{\bloomRF{} Range-Lookup}
\label{alg:rangeq}
\small
	\SetKwFunction{FRangeQ}{RangeLookup}
	\SetKwProg{Pn}{Function}{:}{\KwRet}
	\Pn{\FRangeQ{ \texttt{l\_key, r\_key} \  } }{
		Initialize: counter $i$ with \textbf{|H|}; vector \emph{intervals} = $\varnothing$\;
		\label{alg:rq:line:init} 
		Let: $check \gets$ if interval $[l\_key, r\_key]$ is in layer $i - 1$\;
		\label{alg:rq:line:start:check} 
		Append $check$ to vector $intervals$\;
		\While{$(i \gets i - 1 ) \geq 0 $} {
		\label{alg:rq:line:lower} 
			Let $mask\! \gets\! \textbf{bit\_mask}(check.l\_key, \ check.r\_key)$\;	
			\label{alg:rq:line:mask}	
			$pos \gets {\bf H_{i}}$($check.l\_key$ \!$>\!>$\! {\bf b} $\cdot$ \textit{i}) {\bf mod} 
			\label{alg:rq:line:pos}	
			{\bf SizeOf}(\! filter\! )\; \label{alg:rq:line:pos}
			\If{$(mask \gets (filter[pos] \ \& \ mask)) = 0$} { 
			\label{alg:rq:line:probe} 
				remove $check$ from $intervals$\; 
				\If{$intervals = \varnothing$} {
				\label{alg:rq:line:check:empty} 
					\KwRet{$false$}\;
				}
				\Else{
					$check \gets intervals.$\textbf{next}()\;
					\If{$check.layer < i$} {
					\label{alg:rq:line:child}
						\textbf{continue}\; 
					} 
					\Else {
					\label{alg:rq:line:sibling}
						\textbf{goto}: \ref{alg:rq:line:mask}\;
					}
				}
			}
			\ElseIf { \textbf{popcount}(mask) $>$ threshold} {
			\label{alg:rq:line:count:1bits}
				\KwRet{$true$}\;
			} 
			\Else{
			\label{alg:rq:line:inspect:subintervals}
				Append sub-intervals determined by $mask$ to $intervals$, unless on leaf layer\;
				$check \gets intervals.$\textbf{next}()\;
			}
		}
		\KwRet{$true$}\;
	}
\end{algorithm}

We proceed by iteratively inspecting lower layers (line \ref{alg:rq:line:lower}). Initially, we  compute a \emph{bitmask} (line \ref{alg:rq:line:mask}) for the adjacent dyadic intervals with the trace of a TT indicated by \emph{check}-interval. This is possible since: (i) intervals are encoded as bits by the trace-encoding scheme; and since (ii) PMHF ensure local order in a TT trace. Next, the left-key of the current \emph{check}-interval is hashed by the PMHF on the current DTT layer \emph{i}, to determine the bit-array element corresponding to the trace (line \ref{alg:rq:line:pos}). Eventually, the element is fetched from bit-array and overlaid with the bitmask (bitwise \texttt{AND}, line \ref{alg:rq:line:probe}).

If the \emph{mask} is empty, the search interval is not covered by sub-intervals of \emph{check} and no child TTs need to be inspected. Hence, if the vector of \emph{intervals} is empty (line \ref{alg:rq:line:check:empty}) the search interval does not exist. Otherwise, we proceed with the next \emph{check} interval and process it, while distinguishing between a child (line \ref{alg:rq:line:child}) and sibling \emph{check}-intervals (line \ref{alg:rq:line:sibling}).

However, if the resultant \emph{mask} is nonempty, we count the set bits (\texttt{POPCNT}-population count CPU-instruction), to realize an approximate \emph{early-stop} condition. If the number of adjacent set bits is larger than a \emph{threshold}, i.e. 3, the probability $\varepsilon$ of the sub-intervals being empty is very low, and the search range exists with error $\varepsilon$. Otherwise (line \ref{alg:rq:line:inspect:subintervals}), unless we are on the leaf DTT layer, sub-intervals are added for inspection and the algorithm proceeds with the next interval. 


An important property of the range-lookup algorithm is the close adherence to the dyadic interval scheme (DIS). Consequently, it can exclude potentially irrelevant sub- or sibling-intervals early, reducing computations and bit-array fetches. Consider again the example in Fig. \ref{fig:headpict}.e. As a result of Step 1, sibling interval \emph{$\geq$192} and sub-interval \emph{183$\leq$} can be excluded early, saving computations and three bit-array element fetches of $B_{10}$, $B_{9}$ and $B_{7}$. For the same reason it handles \emph{data skew} well. The algorithm is adaptive as the value of \emph{threshold} can be learned for a given workload. 

\noindent{\bf Point-Lookup.} \bloomRF{} hashes the search \emph{key} consecutively with all PMHF (Algorithm \ref{alg:pointq}), computing the vertical representatives of \emph{key}. For each $MH_{i}$, it firstly computes the bit-array position \emph{pos} (Line \ref{alg:pq:line:pos}) and the \emph{mask} within the trace. Secondly, \bloomRF{} probes the bit-array by fetching the element for \emph{pos}. If a bitwise \texttt{AND} with the position \emph{mask} returns zero the \emph{key} is nonexistent and we stop. Otherwise we proceed with the next $MH_{i}$ as the result may be falsely positive. 
\noindent\texttt{Insight:} A \emph{key} is nonexistent if \emph{at least one} bit position \emph{pos}, $ pos \in \{ MH_{i}(key) \}$ is not set in the bit-array.

\begin{algorithm}[t]
\caption{\bloomRF{}  Point-Lookup}
\label{alg:pointq}
\small
\setcounter{AlgoLine}{0}
\SetKwFunction{FPointQ}{PointLookup}
\SetKwProg{Pn}{Function}{:}{\KwRet}
\Pn{\FPointQ{ $key\ $} }{
	Initialize counter $i$ with \textit{zero}\;
	\While{$(i \gets i + 1) \leq \textbf{|H|}$}{ \label{alg:pq:line:while}
			Let $pos \gets {\bf H_{i}}$($key$ \!$>>$\! {\bf b}) {\bf mod} 
												{\bf SizeOf}( filter )\; \label{alg:pq:line:pos}	
			Let $mask \gets 1 << (key \ \& \ (\ \textbf{|T|} - 1\ ))$\; \label{alg:pq:line:mask}	
			\If{$(filter[pos] \ \& \ mask) = 0$}{ \label{alg:pq:line:probe}
				\KwRet{$false$}\; \label{alg:pq:line:stop}	
			}	
			$key \gets key >> \textbf{h}$\;
	}
	\KwRet{$true$}\; 
} 
\end{algorithm}

\noindent{\bf Insertion.} The \texttt{Insert(key)} operation resembles the  \texttt{Point- Lookup} in Algorithm \ref{alg:pointq}. However, instead of probing and performing an early stop (Lines \ref{alg:pq:line:probe} and \ref{alg:pq:line:stop}, Algorithm \ref{alg:pointq}) \bloomRF{} sets the bit position on all traces applying all $MH_{i}$. This is done by a bitwise \texttt{OR}: $filter[pos] \gets filter[pos] \ {\bf |} \ mask$.

%% file: sections/09_optimizations.tex

 \section{Optimizations}
 \label{sect:optimization}
So far we presented basic \bloomRF{}. Although it is simple and effective, there are several important optimization aspects.

\noindent{\bf Mixed Trace Sizes.}
We now relax the assumption of a single TT trace-size for the whole DTT. In fact, the trace-size must be uniform within a DTT layer, however it may vary across the layers. 
As a result of mixed-trace sizes, any two DTT layers may span different number of dyadic levels.
To this end, we define a \emph{height vector} 
$h=(h_0,h_1,\ldots,h_L) \in \mathbb N^{L+1}$
as a vector of numbers, defining the height of the DTT layers.
The bottom-level of each layer is given by
$
  l_i = \sum_{j=0}^i h_j
  .
$

\begin{figure}[b]
	\centering
  \includegraphics[width=\linewidth]{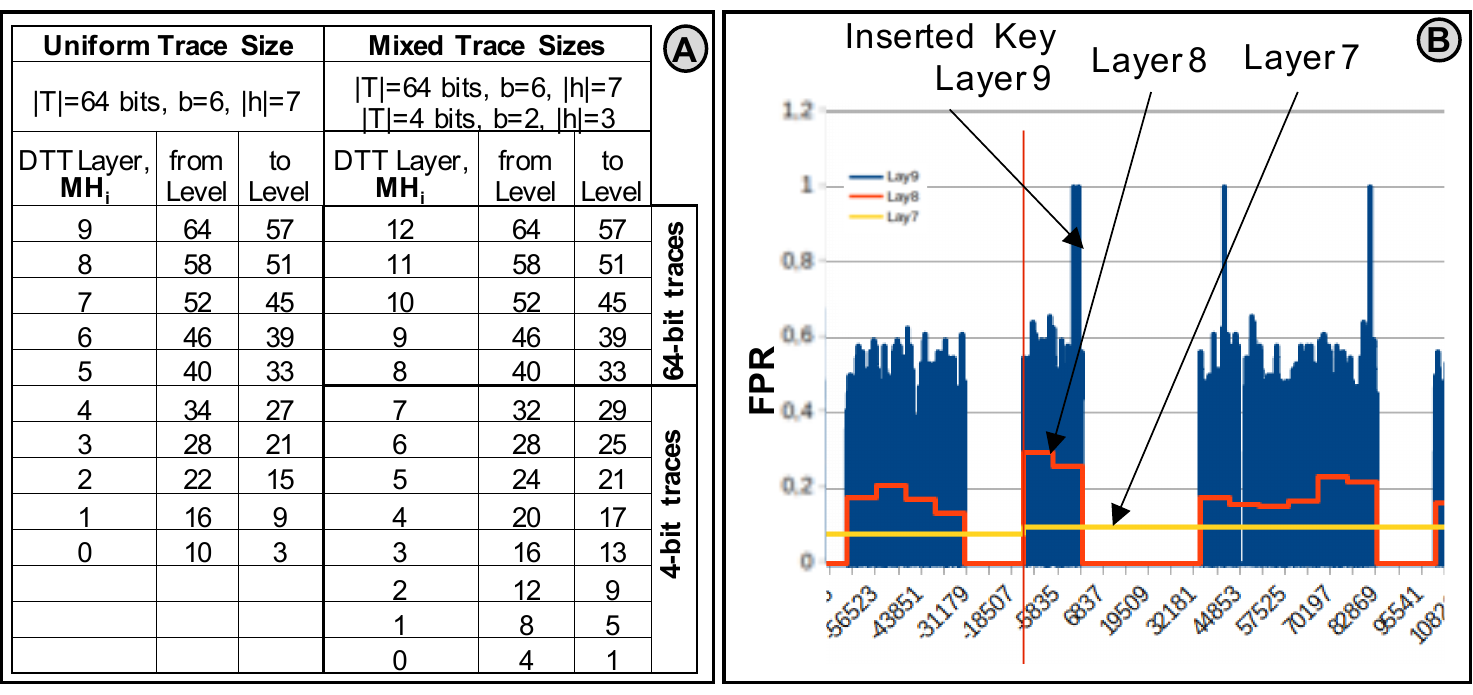}
  \caption{(a) Mixed trace-sizes increase the number of PMHF, and (b)  improve the error-correction.}
  \vspace{-5pt}
  \label{fig:ecc} 	
\end{figure}

Mixed trace-sizes increase the number of PMHF and yield better \emph{error-correction}, better randomization in the Bit-Array and ultimately lower FPR. Smaller trace-sizes $|T|$ imply lower TT height,  larger DTT height 	$|H| = \floor*{  LOG_{(2\cdot |T|)} \left( |D|/|T| \right) }\! +\! 1$ and thus yield more PMHF $MH_{i}(x) ,i\in[1,|H|]$. Consider for example Fig. \ref{fig:ecc}.a. Uniform trace sizes of 64 bits yield a DTT of 10 layers and 10 PMHF. Choosing smaller trace-sizes for the upper DTT layers ($|T|\!=\!4$ bits), and larger sizes for the lower DTT layers increases the number of PMHF to 13. 

\emph{More PMHF yield better vertical error-correction}, which is important for handling skew and long ranges.  To quantify the effect we perform an experiment (Fig. \ref{fig:ecc}.b) where we pick a leaf layer (9) trace containing a data point, point-query all neighboring data points in that trace, all other neighboring layer 9 traces until we leave the parent layer 8 trace, and so forth until we leave the parent layer 7 trace. 
The FPR within the layer 9 trace in question and the direct neighbor is approx. 50\%, since only $MH_{9}(x)$ can be used. (An FPR of 100\% is also possible due to the overlapping of different traces in the same bit-array element.) The avg. FPR on all other neighboring layer 9 traces is 0\% or 50\% since both $MH_{9}(x)$ and $MH_{8}(x)$ are used, while on the layer 8 we observe an avg. FPR of 25\% as  $MH_{9}(x)$, $MH_{8}(x)$ and $MH_{7}(x)$ are used and on layer 7 the FPR sinks to avg. 12.5\%. 

However, more PMHF increase the data volume and reduce the lookup and insert performance as more bits are inserted. Hence the fundamental tradeoff between error-correction, space-requirements and performance impact.

\begin{figure}[t]
	\centering
  \includegraphics[width=\linewidth]{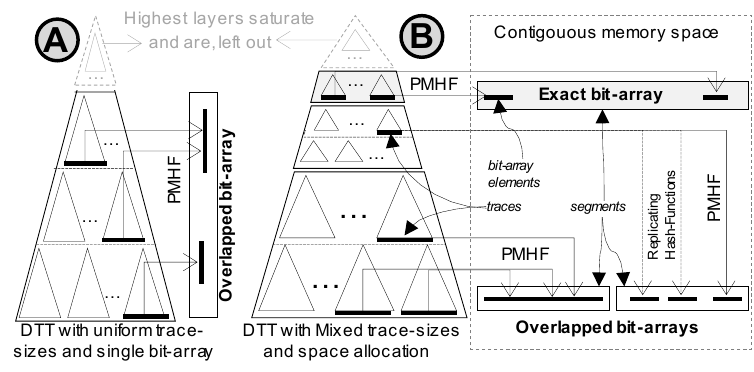}
  \caption{\bloomRFcaption{} space management: (a) basic \bloomRFcaption{} with uniform-sized traces are stored overlapped in a bit-array; and (b) mixed trace-sizes and segment-based space allocation with different error-control policies: exact storage or replicating hashing.
  }
  \vspace{-5pt}
  \label{fig:space:allocation} 	
\end{figure}

\noindent{\bf Replicating Hash-Functions.} 
To control the error-rate in overlapped bit-array, \bloomRF{} may introduce zero or more \emph{replicating hash-functions} ($_{R}$HF) per DTT layer in addition to the PMHF (Fig. \ref{fig:space:allocation}.b). They write replica of the trace, originally positioned by a PMHF, at further bit-array positions. $_{R}$HF preserve the trace-local order defined by the PMHF $MH_{i}(x)$. However, they reduce the error, that bit-flips incur by overlapping other traces in the bit-array, at the cost of higher occupation of bit-array elements and performance. 

\noindent{\bf Space Management.} 
We now revisit the assumption that traces are stored in an implicitly managed bit-array and introduce the \bloomRF{} space management strategy (Fig. \ref{fig:space:allocation}).

The average trace occupation (average number of set bits per trace) on each layer increases along the DTT height due to its logarithmic nature and as a result of hierarchical hashing (Fig. \ref{fig:hashfunc}.c). 
For example, given a uniform distribution and sufficient number data points/keys {\bf n} in basic \bloomRF{}: lower DTT layers tend to be sparser with lower average trace occupation, while mid-upper layers tend to have higher trace occupation, whereas the highest layers saturate as bits overlap multiple times. 

\textsf{Intuition:} 
Mid-upper layer traces are important for long query ranges, since due to the dyadic expansion of large ranges (Sect. \ref{sect:optimization}.4) the probability of probing an interval on a mid-upper DDT layer is high. Hence, the FPR can be improved by minimizing the error-rate of inexact, overlapped storage. 
\bloomRF{} addresses the above by a \emph{space allocation} strategy, which comprises two orthogonal dimensions.
\begin{itemize}[leftmargin=*,noitemsep,nolistsep,nosep]
	\item \emph{Error-control.} \bloomRF{} can partly control the error-rate. Traces can be stored \emph{exactly}  and thus error-free, without any overlapping, but at maximum space costs. Alternatively, \bloomRF{} may store traces in \emph{overlapped} elements at bit-array positions computed by PMHF only. Such strategy is space-efficient, but has the highest error-rates. Finally,  \bloomRF{} employs an intermediary strategy, based on one or more \emph{replicating hash-functions}, for lower error-rates at the cost of more space or higher occupation. 
	\item \emph{Segmentation.}  \bloomRF{} allocates contiguous memory space for the filter. However, it separates it into  exclusive or shared \emph{segments} (Fig. \ref{fig:space:allocation}.b), governed by an error-control policy. \emph{Exclusive} implies that one or more DTT layers are placed in a separate and exclusively reserved part of the memory space. Whereas, \emph{shared} implies that traces of several DTT layers are placed in a common bit-array.
\end{itemize}

\textsf{Summary:} The combined effect of the above optimizations is depicted in Fig. \ref{fig:space:allocation}. To handle large query ranges and large number of keys \bloomRF{} typically employs the following strategy: 
(i) sparser lower DTT layers with large trace sizes (e.g. 64-bit) are packed denser together in a common overlapped bit-array with a single PMHF; 
(ii) mid-layers with small trace sizes (e.g. 8-bit or smaller) are stored in a separate and sparser bit-array with one more replicating hash-functions besides the PMHF to lower the error-rates; 
(iii) a mid-upper DTT layer with small trace sizes (e.g. 8-bit or smaller) is stored separately and exactly; while 
(iv) the highest layers atop are ignored as they saturate. In this sense, basic \bloomRF{} (Fig. \ref{fig:space:allocation}.a) is one possible space configuration, with a single shared segment, based on PMHF, without any $_{R}HF$ or an exact layer.
More formally: in a DTT given by a height vector
$h\in \mathbb N^{L+1}$, the space management strategy mandates that a layer $\ell_0$ is stored exactly.
Layer $\ell_i$, $i \in \{1,2,\ldots, L\}$, is stored with
$k_i$ hash-functions in a bit-array $B=\{1,2,\ldots m\}$:
$
  \varphi_j^{\ell_i}
  \colon
  \mathcal I^{\ell_i} \rightarrow B
  ,
  j \in \{1,2, \ldots, k_i\}
$.

\noindent{\bf Offline optimizations.} 
 Due to their novelty we focus only on online scenarios, i.e. queries being performed, as keys are being inserted. 
However, there are a number of possible offline optimizations that \bloomRF{} can perform if the complete dataset were known \emph{a priori}. Due to its parallel nature \bloomRF{} allows for \emph{bulk-insertions} or fast multi-threaded insertions on a set of keys with some common prefix. Furthermore, the exact layer can be \emph{compressed}  with a lightweight compression scheme.

%% file: sections/07_parameter.tex

\section{\bloomRFtitle{} Parameter Selection}
\label{sect:param}

\noindent{\bf FPR Modeling.}
Particularly with view of the optimizations in the previous section
\bloomRF{} offers a number of internal parameters:
Mixed traces sizes with $L$ layers defined by
a height-vector $h \in \mathbb N^{L+1}$,
replicated hash-functions which define $k_i$ hash-functions for layer $i$
and space management with $M$ segments of sizes $m\in \mathbb N^M$,
where layer $i$ is assigned a segment $j_i$.
We further assume, that the keys are taken from
a uniform distribution and layer 0 is stored exactly.

We introduce a model to estimate the false-positive rate
of the dyadic intervals on the lowest level 
$l_i=\sum_{j=0}^i h_j$ of each layer $i$.
These can be classified as intervals, which according to the filter are: (a) empty ($\tn_i$); or are (b) non-empty and include a key ($\tp_i$);  or are (c) non-empty and do not include a key ($\fp_i$).
Then the false-positive-rate of layer $i$ is
$ \fpr_i = \fp_i/(\fp_i+\tn_i)$.
The number of true-positives on each layer can be estimated
using the distribution of the keys.
For uniform distribution $n$ keys lie in approximately
$n$ dyadic intervals on large enough levels.
The number of true positives can be estimated by the expression: $\tp_i=\min(n,2^{\ell_i})$.

We give recursive formulas for $\fp_i$ and $\tn_i$.
Since layer $\ell_0$ is stored exactly
$\fp_0=0$ and $\tn_0=2^{\ell_0}-tp_0$. 
Suppose we have computed $\fp_i$ and $\tn_i$ of layer $i$.
Each dyadic interval $I$ on level $\ell_i$ includes
$2^{h_i}$ dyadic intervals on level $\ell_{i+1}$.
If $I$ is a true-negative,
then all $2^{h_i}$ intervals are also true-negatives.
If $I$ is a false- or true-positive,
then some of the $2^{h_i}$ intervals can have false-positives.
By our estimate $\tp_{i+1}$ are true-positive. Hence the number of
potentially false positive intervals is
$
 \fp^{\pot}_i
 \!=\!
 2^{h_i}(\fp_i+\tp_i)\!-\!\tp_{i+1}
$
.
Whether such an interval is false positive depends on the fill rate $\beta_i$
of the segment $j_i$ of the bit-array.
The fill rate is the relative frequency that a bit corresponding to
an interval not including a key is still set to 1.
With this we get
$\fp_{i+1}=\beta_i^{k_i} \fp^{\pot}_i$
and
$\tn_{i+1}=2^{h_i} \tn_i + (1-\beta_i^{k_i})\fp^{\pot}_i$
.
The false positive rates of the remaining levels can be estimated
from the computed estimates by combinatorial formulas.

\noindent{\bf  \bloomRFcaption{} Tuning Advisor.}
Given standard parameters such as the number of keys $n$, the memory budget $m$ and considering the query range size, the goal of the tuning advisor is to compute and select an appropriate \bloomRF{} configuration. It comprises the number of segments and their sizes, the height-vector and the exact layer, the trace sizes and the number of replicating hash functions. In the following we describe the procedure. 
First, we estimate the FPR of the dyadic intervals on each level. The goal is to minimize the maximum FPR $\fpr_m$. Since the largest rates result from the higher  levels (= large intervals), small intervals (= lower levels) are under-prioritized. Consequently, we also estimate the point query FPR $\fpr_p$ and minimize $\fpr_w^2=\fpr_m^2 + C^{2}\cdot \fpr_p^2$, where $C>1$ is a constant, since $\fpr_p < \fpr_m$ always holds.
To reduce the number of parameters we apply the heuristic that the exact layer should use less then $60\%$ of the memory $m$. 
Next, we process the next two smaller levels. For a given exact layer we compute a height-vector $h$, the number of hash-functions $k$ and the memory allocation per segment $j$. For example, if level 28 is exact: 
$h=(28,2,2,4,7,7,7,7)$,
$k=(1,2,1,1,1,1,1,1)$
and $j = (1,2,2,2,3,3,3,3)$.
We aim for three segments: $m_1$ for the exact, $m_2$ for the middle and $m_3$ for the lower layers, $m=m_1+m_2+m_3$ (Fig. \ref{fig:space:allocation}). Since $m_1=2^{\ell_0}$ it remains to select $m_2$. Increasing $m_2$ reduces $\fpr_m$, but increases $\fpr_p$. We select $m_2$ such that $\fpr_w$ attains a minimum.
With this procedure the tuning advisor computes a \bloomRF{} configuration for each choice for the exact layer. Finally, the advisor ranks those configurations and selects the one with the smallest $\fpr_w$. If a configuration cannot be computed the advisor resorts to basic \bloomRF{}, which can handle well mid-sized query ranges across a variety of space budgets.
For completeness, Table \ref{tab:bloomRF:config} shows the computed configurations for the experimental evaluation datasets.

\begin{table}[!b]
\small
\centering
\caption{\bloomRFcaption{} Space-allocation configurations for RocksDB and Standalone. Segment sizes in [MB]. Hash-Functions ($k_{i}$) include PMHF and $_{R}HF$.}
\vspace{-10pt}
\begin{tabular}{|r|r|r|r|r|r|r|}
	\hline
	& \multicolumn{3}{c|}{2.06M keys (1GB SST)}& \multicolumn{3}{c|}{50M keys (Standalone)}\\
	Bits/Key & $S_{1}$/$k_{i}$& $S_{2}$/$k_{i}$   & $S_{3}$/$k_{i}$ & $S_{1}$/$k_{i}$ & $S_{2}$/$k_{i}$ & $S_{3}$/$k_{i}$ \\
	\hline
	$16<$ & \multicolumn{6}{c|}{small- and mid-ranges $\rightarrow$ basic \bloomRF{}}   \\                        
	\hline
	\hline
	$\geq$ 16 & \multicolumn{3}{c|}{large ranges}  & \multicolumn{3}{c|}{mid- and large-ranges} \\                        
	\hline
	16	&1/1	&2.10/7		&0.84/5		& 32/1	& 49/4	&14.36/4\\
	\hline
	18	&1/1	&2.55/7	&0.88/5		& 32/1	& 61/4 &14.28/4\\
	\hline
	20	& 2/1	& 2,15/5	& 0.77/5	& 32/1	& 73/4	& 14.20/4\\
	\hline
	22	& 2/1	& 2.60/5	 & 0.82/5	& 64/1	& 52/4	& 15.13/4\\
	\hline
	24	& 2/1	& 3.05/5	& 0.86/5	& 64/1	& 64/4	& 15.05/4\\
	\hline
	26	& 2/1	& 3.50/5	& 0.90/5	& 64/1	& 76/4	& 14.97/4\\
	\hline
\end{tabular}
\label{tab:bloomRF:config}
\end{table}

%% file: sections/06_datatypes.tex

\section{Datatype Support}
\label{sect:datatypes}
 
\noindent{\bf Variable-length strings.} The string support in \bloomRF{} is similar to that of SuRF-Hash \cite{Zhang:SURF:SIGMOD:2018}. \bloomRF{} takes the first seven characters in the seven most-significant bytes and computes a one-byte hash-code for the rest of the variable-length string also considering the length. It is then placed in the least significant byte. This way \bloomRF{} achieves a UINT64 representation of variable length-strings.

\noindent{\bf Floating-Point Numbers.} 
Floating-point numbers are represented with $q$ bits for the mantissa $\mu$,
$r$ bits for the exponent $e$ and one bit for the sign $s$.
For a bit combination $x$ the represented value is
$fl(x) = s \cdot \mu \cdot 2^e$.
The bit combinations $x$ are ordered as binary numbers.
Since floating-point numbers have a sign, this order is reversed
for negative numbers and can therefore not be used for \bloomRF{}.

Instead, we use a map $\varphi$ with
$\varphi(x) = x + 2^{q+r}$ if $x_{q+r}=0$ and
$\varphi(x) = \overline{x}$ (bitwise inverse) otherwise,
which is a monotone coding, i.e.,
$\varphi(x) < \varphi(y) \Leftrightarrow fl(x) < fl(y)$.
For all operations, we use $\varphi(x)$ instead of $x$.
To insert $x$ into \bloomRF{}, we insert $\varphi(x)$.
For a point-query of $x$ we test $\varphi(x)$.
For a range-query $[x,y]$, we perform a range-query
with  $[\varphi(x),\varphi(y)]$.

 \noindent{\bf Multi-Attribute \bloomRFtitle{}.} 
The ability to filter on multiple attributes simultaneously is under-investigated, but necessary for complex operations in interactive analytics, scientific packages, IoT and AI. \bloomRF{} supports two-dimensional filtering with reduced precision.  Hence, the core idea is to \emph{concatenate} the attribute-values and insert them in \emph{both} combinations. For instance, \bloomRF\texttt{(A,B)} will concatenate the values of attributes \texttt{A} and \texttt{B}, and insert them in both combinations \texttt{<A,B>} and \texttt{<B,A>}. The increased space-requirements are lowered by reducing the precision of \texttt{A} and \texttt{B}, e.g.  to a 32-bit integer. As a result \bloomRF{} can answer queries such as \texttt{A<42 \!AND\! B=4711}, \texttt{A=42 \!AND\! B>4711} or \texttt{A=42 \!AND\! B=4711}.

%% file: sections/10_evaluation.tex

\section{Experimental Evaluation}
\label{sect:eval}
\noindent{\bf Integration in RocksDB.} \bloomRF{} has been implemented in a standalone library and has been integrated in RocksDB v6.3.6 by means of a filter policy. For persistence it implements its own serialization/deserialization mechanism. We follow the RocksDB parameter settings described in \cite{Dayan:Rosetta:SIGMOD:2020}.

\noindent{\bf Baselines.} Throughout the evaluation the following baselines are used: \BFs{}, Prefix-\BFs{} and fence pointers as well as state-of-the-art point-range filters such as SuRF \cite{Zhang:SURF:SIGMOD:2018,surf:lib,surf:rocksdb} and Rosetta \cite{Dayan:Rosetta:SIGMOD:2020}. We perform two types of experiments. First, \emph{system-level} experiments, where all baselines are compared in RocksDB v6.3.6 to stress the overall effects in a real system. Second, \emph{standalone} experiments, to stress specific aspects in isolation.

\noindent{\bf Workloads.} Throughout the evaluation we use a set of different workloads. Firstly, we employ a derivative of YCSB \cite{ycsb:2010} Workload E, which is range-scan intensive. The dataset comprises 50M 64-bit integer keys, while the values are 512 bytes long. The data is uniformly distributed, while the workloads are of normal, uniform and Zipfian distributions. We issue $10^{5}$ queries of a single  \emph{fixed} range-size that is specified in the respective experiments. All point- and range-queries in this workload are \emph{empty} (unless specified otherwise), which is the worst-case that also allows reporting the absolute FPR and its workload execution-time implications. 
Rosetta and \bloomRF{} rely on parameter tuning methods that compute the proper filter-configurations, for given space budgets, number of keys and range sizes. SuRF, however, requires a suffix-length parameter setting to tune itself to a space budget and trade off FPR, by selecting the appropriate variant. 
If the respective space budget cannot be achieved we select the next possible one. 
Secondly, for the floating point experiments we use a timeseries dataset from NASA\cite{Kepler:2016}. 
Whereas for the multi-attribute experiments we utilize a dataset from the Sloan Digital Sky Survey DR16 \cite{sloan:SDSS:2019}, comprising integer and float attributes.
For strings we use a Wikipedia Extraction dataset  by DBpedia \cite{dbpedia:2020}. 

\noindent{\bf Experimental Setup.} The experimental server is equipped with an Intel E5-1620 3.50GHz CPU, 32GB DDR4 a 512GB Samsung 850 SSD, a 4TB HDD, and runs Ubuntu 16.04. 

\begin{figure*}[!ht]
	\begin{center}
		\includegraphics[width=\textwidth]{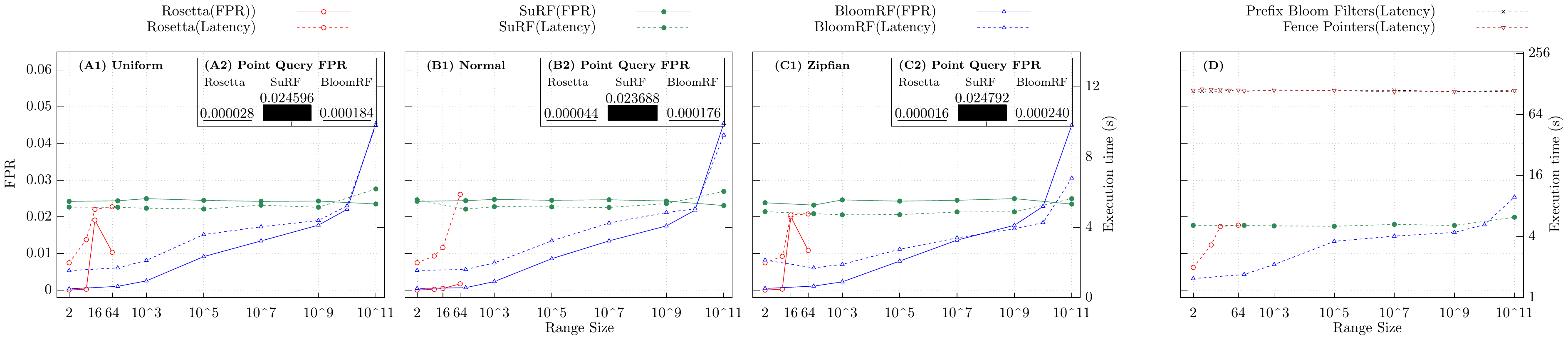}
		\caption{\bloomRFcaption{} has good performance for a variety of ranges and workload distributions in RocksDB.}
		\label{fig:exp1}
	\end{center}
\end{figure*} 

\noindent{\bf Experiment 1: \bloomRFcaption{} is general and can handle various query ranges, from large to small.} 
We begin with a general experiment in RocksDB, comparing \bloomRF{} to SuRF and Rosetta across a variety of query range sizes and workload distributions (Fig. \ref{fig:exp1}.A1, B1 and C1). We employ 50M uniform distributed keys and a space budget of 22 bits/key that is favorable for all approaches. As Rosetta is designed for \emph{small} query ranges $|R|$ it has lower FPR than \bloomRF{} for $|R|\!\leq$ 8 or 16 depending on the distribution, i.e. 0.00023 versus 0.00062 mean FPR. However, \bloomRF{} has 2$\times$ to 3$\times$ lower latency due to its efficient range lookup algorithm, i.e approx. 3.5$\mu$s vs. approx. 7..10.8$\mu$s filter time per seek. With $16\!\leq\!|R|\!\leq64$ \bloomRF{} has better FPR. In comparison, SuRF has a mean FPR of 0.0245 a latency of approx. 5.1$\mu$s.
For \emph{mid-sized} ranges ($10^{2}\!\leq\!|R|\!\leq\!10^{5}$) \bloomRF{} has a mean FPR of 0.001 to 0.009 and a filter time per seek of approx. approx. 3.5$\mu$s, while SuRF retains its mean FPR of 0.0245 and latency of approx. 5.1$\mu$s. 
Because of its trie suffix-truncation techniques SuRF is especially effective for \emph{large query ranges} i.e. ($10^{7}\!\leq\!|R|\!\leq\!10^{11}$).  For $|R|\!\leq\!10^{10}$ \bloomRF{} has 
mean FPR of 0.0105 to 0.0177 and a latency of approx. approx. 3.5$\mu$s, while SuRF retains its mean FPR of 0.0245 and latency of approx. 5.1$\mu$s. However, $|R|\!\geq\!10^{11}$ \bloomRF{}'s FPR increases to 0,0454, while SuRF's FPR falls to 0.0232 with same latencies. (The sudden rise in \bloomRF{} workload latency at $|R|\!=\!10^{11}$ is due to approx. 1\% non-empty ranges generated by the workload driver because of the large interval size.) Hence, \bloomRF{} outperforms Rosetta and SuRF and has better FPR for a wide variety of range sizes and workload distributions, while being competitive for very small and very large ranges.

Under the same settings, we also investigate the FPR of \emph{point queries} (Fig.\ref{fig:exp1}.A2, B2 and C2 shown as figure-in-figure in Fig. \ref{fig:exp1}). Due to Rosetta's design decision and configuration method the lowest filter-layer is very accurate and can answer point-lookups with low FPR. It is therefore not surprising that Rosetta has the lowest point-query FPR. \bloomRF{} needs more space for the DTT layers, which allows for querying longer ranges, but yields slightly higher point FPR. Because of the trie truncation SuRF has the highest point FPR.

Compared against Prefix-\BFs{} and fence pointers (Fig.\ref{fig:exp1}.D) all point-range-filters excel in terms of performance. 

\textsf{Insight:} \bloomRF{} covers a broad set of query ranges and out performs all baselines, except for very small and very long ranges.

\begin{figure*}[!t]
	\begin{center}
		\includegraphics[width=\textwidth]{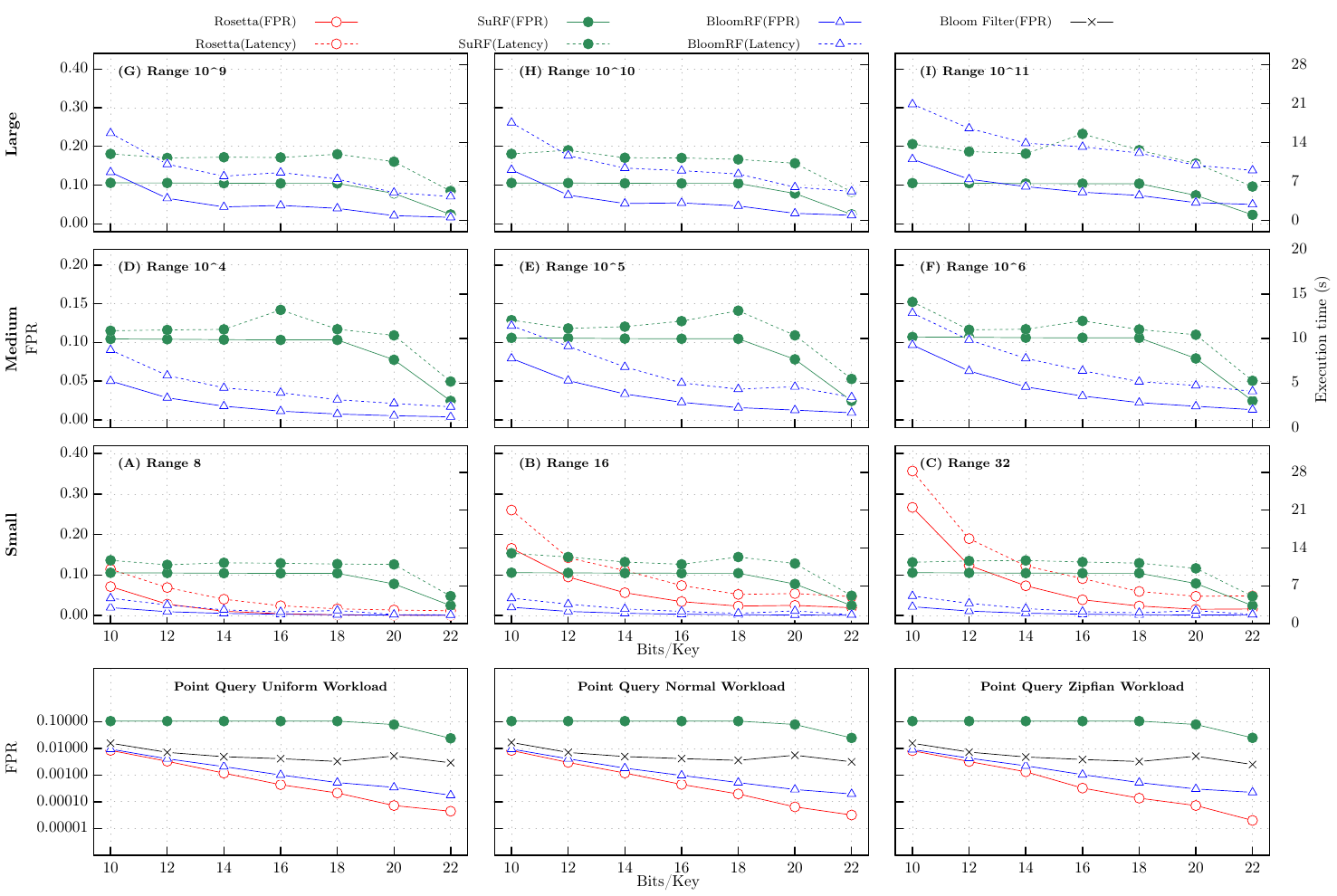}
		\caption{\bloomRFcaption{} is efficient, with better performance for different space budgets and query ranges in RocksDB.}
		\label{fig:exp2}
	\end{center}
\end{figure*}

\noindent{\bf Experiment 2: \bloomRFcaption{} is efficient.} 
In the next experiment (Fig.\ref{fig:exp2}) we continue our comparison, by varying the space budget (bits per key) in RocksDB. We start from the 22 bits/key (favorable for all approaches) that were used in the previous experiment and proceed to 10 bits/key, which is typical for standard \BFs{}. As we go, range lookups with different query range sizes (not mixed) are performed. We use 50M keys; data and workload are uniformly distributed.

\bloomRF{} outperforms or is competitive to Rosetta for \emph{small} ranges like 8, 16, 32 (Fig. \ref{fig:exp2}.A-C) and has similar or better FPR. Although, the performance and FPR of SuRF improve significantly with $\geq\!18$ bits/key for \emph{mid-sized} query ranges such as $10^{4}..10^{6}$, \bloomRF{} outperforms it and improves with increasing space budgets (Fig. \ref{fig:exp2}.D-F). 

For \emph{long} query ranges such as $10^{9}..10^{10}$ (Fig. \ref{fig:exp2}.G-I), in RocksDB \bloomRF{} outperforms SuRF 1.3$\times$ and achieves lower FPR. However, with very long ranges $|R|\!\geq\!10^{11}$ SuRF naturally achieves excellent FPR  0.023 due to the trie-encoding, while \bloomRF{}'s FPR is 0.05. The sudden rise in workload latency at 22 bits/key is due to approx. 1\% non-empty ranges generated by the workload driver due to the large range size. 

For \emph{point-lookups} (Fig. \ref{fig:exp2}, bottom row) not surprisingly Rosetta achieves better FPR than \bloomRF{}. Both are more accurate than the RocksDB \BF{}. In terms of throughput \bloomRF{} outperforms Rosetta 7\% to 44\% at 10 and 22 bits/key, respectively (under uniform workloads). 

\textsf{Insight:} Considering the performance and FPR of 
\bloomRF{} at smaller space budgets (Fig \ref{fig:exp2}), i.e. $\leq\!18$ bits/key, we observe that \bloomRF{} is efficient  and offers good: (i) performance per bits/key; and (ii) FPR per bits/key.

\begin{figure}[!b]
    \includegraphics[width=1.03\columnwidth]{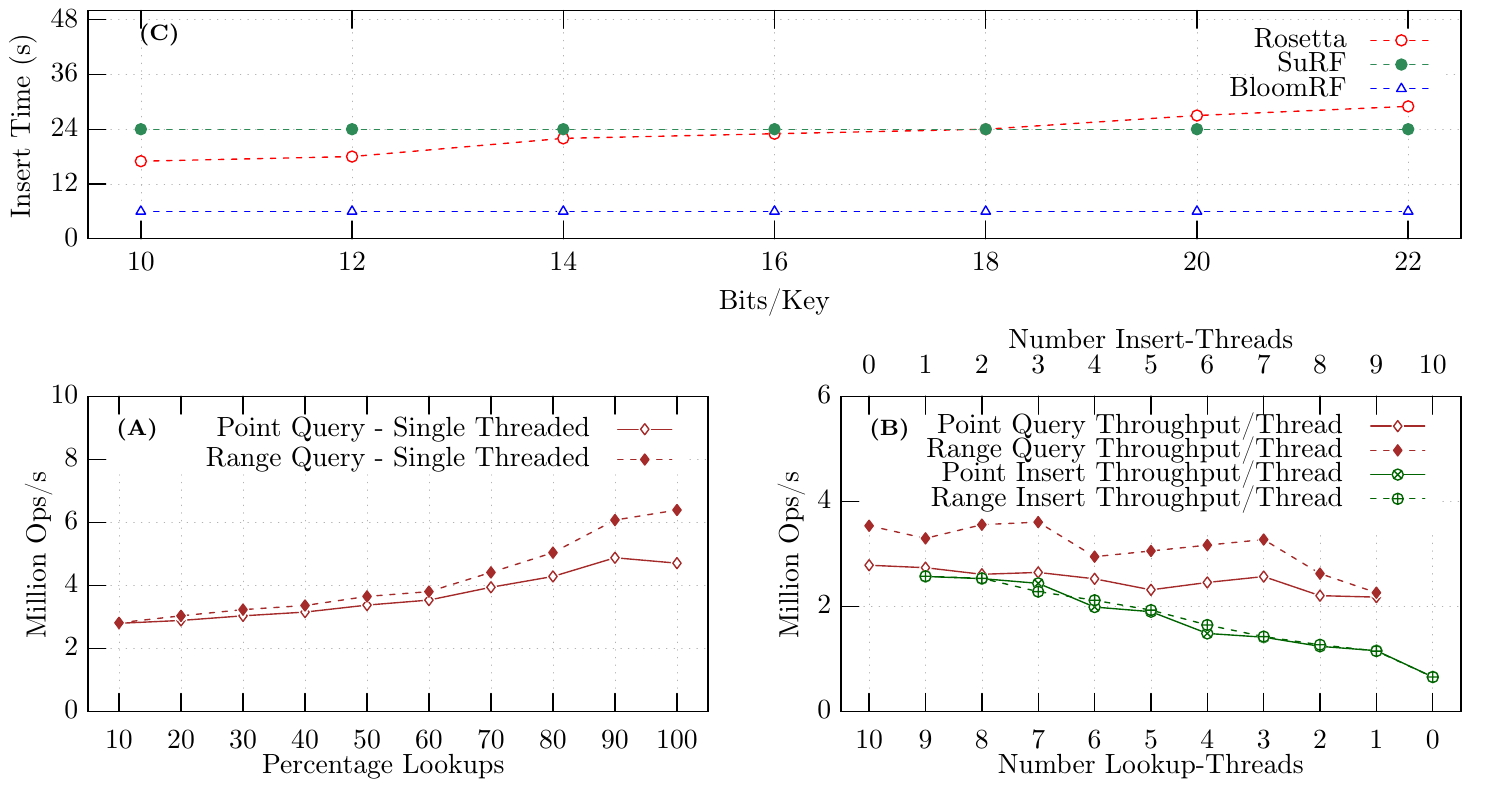}
    \caption{Impact of insertions on probe performance.}
    \vspace{-5pt}
    \label{fig:exp3} 	
\end{figure}

\noindent{\bf Experiment 3: \bloomRFcaption{} is online and concurrent insertions have acceptable impact on its probe-performance.} 
\bloomRF{} is \emph{online} and can perform queries, while keys are being simultaneously inserted. We now investigate the impact of concurrent insertions on query performance, by inserting 50M keys with different insert/lookup ratios in a standalone setting. Clearly, insert-performance is lower than lookup-performance. In single-threaded settings (Fig \ref{fig:exp3}.A), the overall throughput increases with higher lookup ratios. Hence, the impact of concurrent insertions is acceptable. 
A deeper analysis in multi-threaded settings (Fig \ref{fig:exp3}.B) with varying the number of concurrent lookup/insertion-threads shows that concurrent insertions have marginal impact on the lookup performance per thread, as it stays relatively constant. The overall insert-throughput increases with more threads, although the throughput per insert-thread decreases. Furthermore, \bloomRF{} does not require the entire dataset or any preparatory steps, e.g. sorting, prior to serving queries. Together with the good query performance this yields low filter-construction costs (Fig \ref{fig:exp3}.C).

\textsf{Insight:} \bloomRF{} is online and has good performance with different insert/lookup mixes and low construction costs.

\begin{figure}[!t]
    \includegraphics[width=\columnwidth]{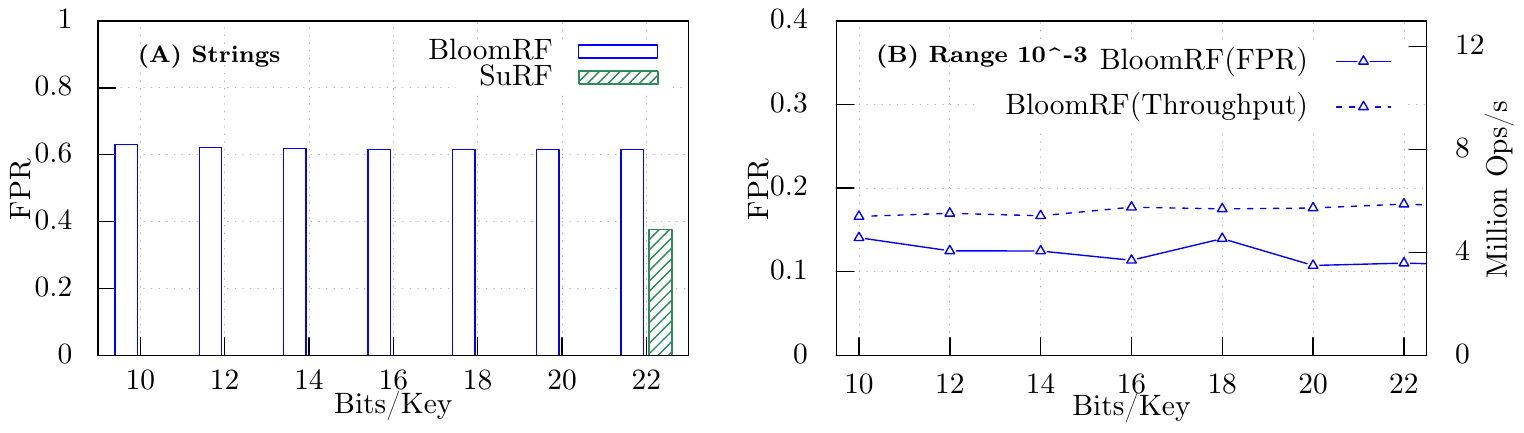}
    \caption{\bloomRFcaption{} with (a) strings and (b) floats.}
    \vspace{-5pt}
    \label{fig:exp4} 	
\end{figure}

\noindent{\bf Experiment 4: \bloomRFcaption{} can handle strings and floats.} 
To investigate variable-length string performance, we use a 1 million key Wikipedia Extraction dataset by DBpedia \cite{dbpedia:2020}. In particular we import the \emph{abstracts} portion. In a standalone setting, we issue 1 million uniformly distributed  queries with a range size of 128. The results are shown in Fig. \ref{fig:exp4}.a. SuRF's trie-encoding can optimally cover strings with common prefixes, yielding very good FPR. The small query range is suboptimal for the string encoding of \bloomRF{} (Sect. \ref{sect:datatypes}). Nonetheless, it achieves an FPR of 0.61 to 0.63 depending on the space budget.

For floating-point numbers we rely on a public dataset from NASA Kepler mission \cite{Kepler:2016}, containing positive and negative numbers. In a standalone setting, we execute 1.8 million range queries, with range sizes of $10^{-3}$. As to the best of our knowledge there are no other baselines we only show the performance of \bloomRF{} (Fig. \ref{fig:exp4}.b). The FPR ranges between approx. 0.14 and 0.1.

\textsf{Insight:} \bloomRF{} can handle floats besides integers and strings, which is relevant for scientific DBMS or AI.

\noindent{\bf Experiment 5: \bloomRFcaption{} can serve as multi-attribute filter.} 
We evaluate multi-attribute querying in \bloomRF{} on a Sloan Digital Sky Survey DR16 \cite{sloan:SDSS:2019} and extract the \emph{ObjectID} and the \emph{Run} columns. Their values roughly follow a normal distribution. In a standalone setting, we compare a multi-attribute \bloomRFparam{Run,ObjectID} against two separate \bloomRFparam{Run} and \bloomRFparam{ObjectID}. Against \bloomRFparam{Run,ObjectID} we execute the combined conjunctive condition:
\texttt{Run<300 AND ObjectID=Const}. As baseline we execute the two sub-terms against separate \bloomRF{}s, combining the probe-results conjunctively, i.e. \texttt{Run<300} is executed against \bloomRFparam{Run}, while \texttt{ObjectID=Const} is checked with \bloomRFparam{ObjectID}. 

\begin{figure}[!b]
    \includegraphics[width=\columnwidth]{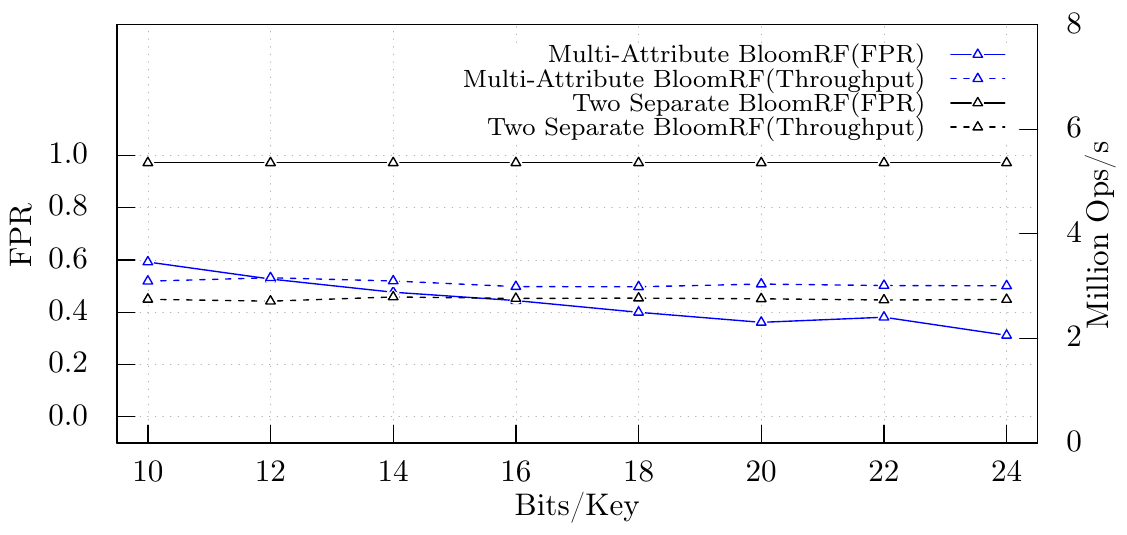}
    \caption{Dual-Attribute \bloomRFcaption{} outperforms two separate \bloomRFcaption{}s on individual attributes.}
    \vspace{-5pt}
    \label{fig:exp5} 	
\end{figure}

As shown in Fig. \ref{fig:exp5} \bloomRFparam{Run,ObjectID} yields better FPR than the combined FPR of the two separate filter-lookups \bloomRFparam{Run} and \bloomRFparam{ObjectID}. This observation is surprising since the separate filters operate on 64-bit integers, while the multi-attribute \bloomRF{} reduces precision (Sect. \ref{sect:datatypes}) and operates on 32-bit integers. The experiment is bound to the combinatorial space, as to the best of our knowledge it is only alternative with single-attribute filters. Although the two separate filter-lookups are executed concurrently, the multi-attribute filter outperforms them and has lower latency.

%% file: sections/11_relWork.tex

\section{Related Work}
\label{sect:relwork}

The \emph{Adaptive Range Filter (ARF)} \cite{Alexiou:ARF:VLDB:2013} is one of the first approaches to \emph{indirectly} describe the use simple form of dyadic numbering scheme to compute the covering intervals of a point. ARF, however, relies on a binary tree as a data structure and a powerful set of (learning) optimizations. Furthermore, like \bloomRF{}, ARF relies on the concept of covering the whole domain of the datatype. \emph{SuRF} \cite{Zhang:SURF:SIGMOD:2018} shows the full potential of trie-based filters (Fast Succinct Trie) with a powerful encoding scheme (LOUDS-Dense/Sparse). \bloomRF{} also relies on an (TT) encoding scheme, however it takes a different research avenue and relies on the dyadic interval scheme.

\emph{Dyadic intervals} and \emph{dyadic decomposition} of ranges have been first utilized point-range-filtering by 
\emph{Rosetta} \cite{Dayan:Rosetta:SIGMOD:2020}. However, the concept has been applied to a wider range of other applications such as stream processing and summarization \cite{Cormode:CountMinSketch:JA:2005}, hot/cold data separation techniques \cite{Cormode:HotCold:TODS:2005} or persistent sketches \cite{Peng:persistentBF:SIGMOD:2018} as pointed out by \cite{Dayan:Rosetta:SIGMOD:2020} .  
The \emph{Segment Trees} employed by  \cite{Dayan:Rosetta:SIGMOD:2020,Peng:persistentBF:SIGMOD:2018,Cormode:CountMinSketch:JA:2005} resemble the Dyadic Trace-Trees in their dyadic nature. However, DTTs differ in that they: (a) have a virtual root; (b)  have \emph{Trace-Trees} as nodes; where (c) the \emph{TT-encoding scheme} is applied to encode $h$ dyadic levels in a single bit. Another major difference to \cite{Dayan:Rosetta:SIGMOD:2020,Peng:persistentBF:SIGMOD:2018,Cormode:CountMinSketch:JA:2005} is that \bloomRF{} employs piecewise-monotone hash-functions to preserve local order within a Trace-Tree. Both are elementary for range querying. Last but not least, \bloomRF{} is a single data structure, while Rosetta \cite{Dayan:Rosetta:SIGMOD:2020} and PBF \cite{Peng:persistentBF:SIGMOD:2018} rely on a series of \BFs{}.

%% file: sections/12_conclusion.tex

\section{Conclusions}
\label{sect:conclusions}
In this paper we introduce \bloomRF{} as a unified point-range-filter that extends  \BFs{} with range-lookups and can effectively replace them. The core  intuition is to use implicit Dyadic Trace-Trees to represent the set of dyadic intervals covering a data point and a Trace-Tree encoding to encode them efficiently in a compact bit representation. Range querying is supported by piecewise-monotone hash functions. \bloomRF{} supports small to long range sizes. It is online, efficient and auto-tunable.
\\

\noindent{\bf Acknowledgments.} \small
We are deeply grateful to the authors of \cite{Dayan:Rosetta:SIGMOD:2020} and \cite{Zhang:SURF:SIGMOD:2018} for providing/open sourcing the source code of Rosetta/SuRF and thus making the comparative evaluation possible. 
This work has been partially supported by \emph{DFG Grant neoDBMS -- 419942270}; \emph{KPK Services Computing}, \emph{MWK, Baden-W\"urrtemberg, Germany}.